\documentclass[12pt,a4paper,leqno]{amsart}
\usepackage{latexsym, url, color, graphicx, epsfig, float} 
\usepackage[mathscr]{euscript}

\usepackage[bookmarks,bookmarksopen,colorlinks]{hyperref}
\usepackage{thumbpdf}  
\theoremstyle{plain}
\newtheorem{Theorem}{Theorem}[section]
\newtheorem{Definition}[Theorem]{Definition}

\newtheorem{Remark}[Theorem]{Remark}

 \makeatletter
 \newif\ifmsbmloaded@

\input{scrload}

\title{Cryptanalysis of $\mathcal{HFE}$}
\author{\href{mailto:toli@posso.dm.unipi.it}{\texttt{Ilia Toli}}}
\address{{Di\-par\-ti\-men\-to di Matematica
  {\it Leonida Tonelli}\\ via F. Buonarroti 2\\ 56127 Pisa\\
  Italy.\\\href{mailto:toli@posso.dm.unipi.it}
  {\texttt{\tt toli@posso.dm.unipi.it}}}}
\begin{document}



\begin{abstract} I transform the trapdoor problem of $\mathcal{HFE}$ into
  a linear algebra problem.
\end{abstract}

\maketitle

\section{Introduction}

The problem of solving systems of multivariate polynomial equations is
a well-known hard problem. In complexity theory, it is well-known to
be an $\mathcal{NP}$-complete problem. Furthermore, even if we limit
ourselves to the problem of solving systems of multivariate polynomial
of degree two equations, we have again an $\mathcal{NP}$-complete
problem. Therefore, it has been paid a lot of attention, since the
invention of the idea of the $\mathcal{PK}$ cryptography, by Diffie
and Hellman \cite{pkc}.  

A lot of cryptosystems have been proposed since then, where an
eavesdropper is asked to accomplish the hard task of solving systems
of quadratic equations. However, most of them had short lives. The
information that an eavesdropper had on the shape of the private key
usually sufficed to compromise the security.
Some of their cryptanalyses aimed to recover the private key, or something
equivalent, in the sense that gives the same privileges. Other
cryptanalyses reduce the problem to accessible exhaustive searches,
and so on. Recall that the ultimate task of the cryptanalysis is
recovering cleartexts rather than recovering meticulously the whole set
of the numberlets of the $\mathcal{PK}$ \cite{COURTTM}.

In this paper we focus on $\mathcal{HFE}$. It is a
$\mathcal{PK}$ cryptosystem first proposed by Patarin
\cite{patarin96hidden}. It is one of the modifications of a
cryptosystem first proposed by Imai and Matsumoto \cite{imai1}, after
having successfully cryptanalyzed it. 

In its main version, its $\mathcal{PK}$ is a system of $n$ 
quadratic polynomial equations in $n$ variables with coefficients in a
finite field $\mathbb{F}_q$, practically $\mathbb{F}_2$. Its private
key is:

\begin{itemize}

\item a basis, up to an isomorphism, of an overfield
$\mathbb{K}\supset\mathbb{F}_q$, $[\mathbb{K}:\mathbb{F}_q] = n$, as an
$\mathbb{F}_q$-vector space;

\item a single univariate polynomial $f$ of a certain form, with
  coefficients in $\mathbb{K}$;

\item two invertible affine transformations of $\mathbb{K}$.

\end{itemize}

Practically, $p=q=2$. However, cryptosystems can be set up for any
choice of $p$, $q$. For simplicity, hereon we assume only
that $p=q$. The other case can be treated almost identically.

In the our cryptanalysis, we find another sparse univariate
polynomial, such that its knowledge reduces eavesdropping to the task
of solving a single univariate polynomial equation. Its solving in
general is an $\mathcal{NP}$-complete problem. Due to its further
structure, in the case of $\mathcal{HFE}$ its solving is a pure linear
algebra matter.

We call the single polynomial that we find in public {\it an alias of
  the  $\mathcal{PK}$}. All of 
the task of recovering it can be performed within $\mathcal{O}(n^6)$ bit
operations. Recall that $n$ is
actually the only security parameter to the legitimate user, and that
the trapdoor problem is subexponential in it.

We assume that the reader is already familiar with $\mathcal{HFE}$.

Most of the symbolic manipulations throughout this paper are done by
means of Singular, Macaulay2, and CoCoA. If there ever are any calculus
mistakes, it is because of the little part done by hand. In any case,
the calculus errors in the examples do not prejudice the algorithms
they illustrate.  

\section{The Cryptosystem}

Let the parties committed to the tasks be:

\begin{itemize}

\item Alice who wants to receive secure messages;

\item Bob who wants to send her secure messages;

\item Eve, the eavesdropper.

\end{itemize}

Alice chooses two finite fields
 $\mathbb{F}_q <\mathbb{K}$,  
 and a basis $\beta_1, \beta_2,\ldots, \beta_n $  of
 $\mathbb{K}$ as an  $\mathbb{F}_q$-vector space. In practice,
 $q=2$. However, it can be any $p^r$, for any $p$ prime, and any
 $r\in\mathbb{N}$. 

Next she takes a univariate polynomial of the form:

\begin{equation}
f(x) = \sum_{i,j} \gamma_{ij}x^{q^{\theta_{ij}} +
  q^{\varphi_{ij}}} + \sum_i\alpha_ix^{q^{\xi_i}}+\mu_0,\label{f}
\end{equation}

\noindent with coefficients in $\mathbb{K}$, and two affine transformations:
$\mathcal{S}$, $\mathcal{T}:$ $\mathbb{K}\rightarrow \mathbb{K}$; one
left, one right. Let $\partial_f$ be the degree (private data) of $f(x)$.

With manipulations that we skip in order to save space, she generates
her $\mathcal{PK}$; a set of $n$ quadratic polynomials of degree two, in
$n$ variables. The interested reader may find details in \cite{imai1,imai2,
  hfe, hpe}. 

Her private key is:

\begin{itemize}

\item the basis $B$ of $\mathbb{K}$ as an $\mathbb{F}_q$-vector space;

\item $\mathcal{S}$, $f$, $\mathcal{T}$.

\end{itemize}

\section{The Cryptanalysis}

Applying invertible affine trans\-for\-ma\-tions is equiv\-a\-lent to
com\-pos\-ing with permutation af\-fi\-ne polynomials. So, Eve knows that
$\mathcal{S}\circ f\circ\mathcal{T}$ in $\mathbb{K}[x]$ 
is a certain univariate polynomial of the same form like (\ref{f}),
but generally of an enormous degree. This is easily seen if one
observes the general form of such a compositum.  So, $\mathcal{S}\circ
f\circ\mathcal{T}$ is rather sparse, too.  

Let $\mathcal{S}$, $\mathcal{T}$ denote the affine polynomials
corresponding to the respective affine transformations, too. Eve can
represent: 

\begin{equation} \quad \mathcal{T}=t+t_0x^{p^0}+\dots
t_{n-1}x^{p^{n-1}}=(x+t)\circ (t_0x^{p^0}+\dots
+t_{n-1}x^{p^{n-1}}). \end{equation}

Next, Eve knows that $\mathcal{S}$ is a permutation polynomial. So, it 
has a single root. Let it be $s'$. So, we have:

\begin{equation}\quad \mathcal{S}=s+s_0x^{p^0}+\dots
  +s_{n-1}x^{p^{n-1}}=(s_0x^{p^0}+\dots 
+s_{n-1}x^{p^{n-1}})\circ (x-s').\end{equation}

So, Eve can think of $\mathcal{S}\circ f\circ\mathcal{T}$ to be of the form:
$$(s_0x^{p^0}+\dots +s_{n-1}x^{p^{n-1}})\circ (x-s')\circ f\circ
(x+t)\circ (t_0x^{p^0}+\dots +t_{n-1}x^{p^{n-1}}).$$

It is easily seen that the polynomial $F=(x+s')\circ f\circ (x+t)$ is
another polynomial of the same shape and degree of $f$. So, Eve may
assume that the transformations $\mathcal{S}$ and $\mathcal{T}$ really
are linear rather than affine, and that the private polynomial is a certain
$F$. She can omit the translations without any loss. 

Let Eve fix the canonical basis of $\mathbb{K}$, or a basis at her
choice, too. She may assume to apply a nondegenerate linear
transformation $\mathcal{L}$ (that she does not know, but she need
not) to the private basis $B$ of $\mathbb{K}$, and to
$\mathcal{S}\circ f\circ\mathcal{T}$ in $\mathbb{K}[x]$. So, she
obtains the canonical basis $I$ of $\mathbb{K}$, and another
univariate pseudoquadratic polynomial $\mathcal{A}=\mathcal{S}\circ 
f\circ\mathcal{T}\circ \mathcal{L}$. As $\mathcal{T}\circ \mathcal{L}$
is just another linear transformation, Eve can assume that she knows
the basis, and the polynomial is of the form $\mathcal{S}\circ 
f\circ\mathcal{T}$.

\begin{Definition} The Hamming weight of a univariate polynomial is the maximum
  of the Hamming weights of the exponents of the monomials with
  nonzero coefficients. \end{Definition}

In order to calculate this single univariate public polynomial
$\mathcal{S}\circ  f\circ\mathcal{T}$, Eve
writes down the pseudoquadratic polynomial of degree at most $q^{n}-1$
in its general form:

\begin{equation}
A_dx^d+A_{d-1}x^{d-1}+\dots +A_1x+A_0,
\end{equation}

\noindent where she considers the $A_i$ like variables. She includes
in such a polynomial only monomials which's exponents have Hamming
weight at most two. So, her number of variables is at most $\frac{n^2+n}{2}+n$.

Next, Eve has to do at most $\frac{n^2+n}{2}+n$ evaluations to the
$\mathcal{PK}$. So, she obtains a linear system of at most
$\frac{n^2+n}{2}+n$ equations in the $\frac{n^2+n}{2}+n$ variables 
$A_i$. Solving it in $\mathbb{K}$ enables Eve to recover
$\mathcal{A}=\mathcal{S}\circ f \circ \mathcal{T}$ in the form
of a univariate polynomial with coefficients in $\mathbb{K}$. It is a
public knowledge that $\mathcal{A}(x)$ exists, and is unique. So, we
expect that the $\frac{n^2+n}{2}+n$ evaluations are necessary and sufficient.

Now Eve has reduced eavesdropping problem to the problem of solving a
single univariate polynomial equation of a certain form and structure
within its field of coefficients. She
possesses the private key, indeed an alias of its. The only problem to
Eve is that such a polynomial generally is of a huge degree. However,
Eve knows that it is isomorphic to a very low degree polynomial of a
certain form.

\begin{Definition} Two polynomials $a(x)$, $b(x)\in F_q[x]/(x^q-x)$
  are called isomorphic iff there exists a permutation polynomial
  $c(x)$ such that $a(x)=b\circ c(x)$ mod $(x^q-x)$ or $a(x)=c\circ
  b(x)$ mod $(x^q-x)$.\end{Definition}

It is obvious that the above definition sets up an equivalence
relation in the ring $F_q[x]/(x^q-x)$.

Let us now split both the private and the public polynomials $f$ and
$\mathcal{A}$ into
three pieces: the quadratic part, the linear part, and the constant
term. Our important observation here is that the linear
transformations sends the quadratic part of $f$ into the quadratic part
of $\mathcal{A}$; the linear part of $f$ into the linear part of
$\mathcal{A}$, and the constant term of $f$ to the constant term of
$\mathcal{A}$ isomorphically and separately; without stirring the
parts. Besides, none of the transformations changes the constant
part. So, we assume the constant part of the polynomial we are looking
for to be known. 

The quadratic and linear part of $\mathcal{A}$ define
respectively a quadratic and a linear form: $Q,L:\mathbb{K}^n
\longrightarrow \mathbb{K}$. The compositions of $f$ with the matrices
$\mathcal{S}$, $\mathcal{T}$ bring the respective forms into new
forms. So, such compositions correspond to joint bases change for
these forms.

The important observations of the paragraph above will help us to
bring the trapdoor problem into a pure linear algebra problem.

Let us get rid for now of the constant part, and pick it up for
last. 

 We write down their matrices. In characteristic two, the
formula that associates a symmetric bilinear form (i.e., a symmetric
matrix) to the quadratic form is: 

\begin{equation}b(x,y)=q(x+y)+q(x)+q(y).\end{equation}

Alice has limitations on the degree $d$ of $f$. Indeed, if it is too
big, the number of the undesired solutions grows a lot. Besides, if
she goes far away with $d$, the problem becomes hard to her, too. In
any case, all
what we are looking for, is to render Eve's position as good as Alice's.

Therefore, we know that the matrix of the quadratic form of the public
polynomial has a tiny rank. We bring it into the canonical form. This
process corresponds to a basis change of $\mathbb{K}^n$ for $\mathcal{A}$. We
apply the same basis change to the matrix of the linear form.

Up to now, we have obtained two matrices, transpose, (those that bring the
quadratic polynomial into the canonical form), and a polynomial. The
polynomial is the sum of the associated polynomials to the matrices of
the new quadratic and linear forms, and of the constant part. This
polynomial has exactly as many 
quadratic monomials as $f$ does. The problem now are the monomials of
the linear part. They generally still preserve their huge degree.

Well, it is a public knowledge that we can apply to the new linear
form matrix a basis change that brings most of its into the canonical
form, apart the minor of its in the same position with the nonzero
minor of the new quadratic matrix. Doing so, and applying the same
basis change to the quadratic matrix, we get rid of most of the linear
monomials, too, and do not cause any change to the quadratic
polynomial. The new polynomial associated to the new quadratic and
linear matrices is more or less of the same degree as $f$. So, Eve is
able to solve it. 

So, she has a
polynomial and two matrices that indeed put her in the same position
with Alice in decryption. This completes breaking $\mathcal{HFE}$.

\begin{Remark}The matrices that Eve finds are with coefficients in
  $\mathbb{K}$. Those of Alice instead, with coefficients in
  $\mathbb{F}$. This is not any sort of problem. Besides, it is a
  public knowledge that Eve should as well limit herself to
  transformations with coefficients in $\mathbb{F}$, and obtain an
  alias of the key, anyway. In practice, there is no reason to do
  so. \end{Remark} 

\subsection{} 

For most of the rest of this paper we give a step-by-step
example of how do we practically recover $\mathcal{A}(x)$, and then
on how do we actually choose well a pair of linear transformations
that enable us to solve it.

\section{A Toy Example}

We are given the following toy $\mathcal{PK}$ from Wolf \cite{wolf2}:

\begin{equation}
\left\{\begin{array}{l}
x_1+x_3+x_1x_2+x_1x_3+x_2x_3\\
x_3+x_1x_3+x_2x_3\\
x_1+x_2+x_3+x_1x_2+x_2x_3+1.\\
\end{array}\right.
\end{equation}

All what we know besides the $\mathcal{PK}$ equations, is that the
base field is $\mathbb{F}_2$, and that the degree of field extension
is {\it 3}. In some fashion, we will have these data public. Without
them, Bob will be unable to encrypt.

We fix the basis $t^2,t,1$ of
$\mathbb{K}=\mathbb{F}_{2^3}$ as an $\mathbb{F}_2$-vector space. We
choose it at
our pleasure. We take $\mathbb{K}=\mathbb{F}_2[t]/(t^3+t+1)$. Again, we
choose the irreducible polynomial of degree $n$ from $\mathbb{F}_2[t]$ for
generating $\mathbb{K}$ at our pleasure.

Now we write the general form of the polynomial we are looking for; an
alias of the private polynomial $f$. It has at most $3^2=9$
terms. 

Explicitely, in this case it is of the form:

\begin{equation}
a+bx+cx^2+dx^3+ex^4+fx^5+gx^6.
\end{equation}

Now we evaluate the $\mathcal{PK}$ in {\it 7} points:
$x=0,1,t,t+1,t^2,t^2+1,t^2+t$. 

The toy values of the parameters render the wrong idea that we will
have to evaluate a generic-coefficients polynomial in the whole set of
the elements of 
the overfield. Indeed, it is very far from being like that. We
need only $n^2$ evaluations. Card $\mathbb{K}=p^n$, instead. 

From the evaluations we obtain the following system:

\begin{equation*}
\left\{\begin{array}{l}
a=1\\
a+b+c+d+e+f+g=t^2\\
a+tb+t^2c+(t+1)d+(t^2+t)e+(t^2+t+1)f+(t^2+1)g=0\\
a+(t+1)b+(t^2+1)c+t^2d+(t^2+t+1)e+tf+(t^2+1)g=0\\
a+t^2b+(t^2+t)c+(t^2+1)d+te+(t+1)f+(t^2+t+1)g=t^2\\
a+(t^2+1)b+(t^2+t+1)c+(t^2+t)d+(t+1)e+t^2f+tg=t^2+1\\
a+(t^2+t)b+tc+(t^2+t+1)d+t^2e+(t^2+1)f+(t+1)g=1.
\end{array}\right.
\end{equation*}

We solve this system, and find the our alias key:

\begin{equation}
\mathcal{A}(x)=t^2x^6+(t^2+1)x^5+(t^2+t+1)x^4+(t^2+1)x^3+(t^2+t)x^2+1.
\end{equation}

As the polynomial we are looking for is unique, the solution to the
system above exists, and is unique.
Now Eve has only to solve the equation $\mathcal{A}(x)=y$ in order to
recover $x$. Even though it is of an enormous degree, the number of
solutions that Eve finds is equal to those that Alice is expected to
find. This is a public knowledge. Eve, too, 
can descard undesired solutions by the same means that Alice
does. Much the same like Alice. The last task for Eve is that of
recovering two suitable matrices that lower the degree of $\mathcal{A}(x)$.

\section{Conclusions}

\subsection{}

In $\mathcal{HFE}$ the $\mathcal{PK}$ hides a single univariate
pseudoquadratic polynomial. In any fashion, this polynomial is very
sparse. It has at most $\frac{n^2+n}{2}+n$ terms of a certain well-known
shape. So, in any case, Eve can recover it in $\mathcal{O}(n^6)$ bit
operations, for $n$ the degree of the field extension. Recall that $n$
is Alice's only security parameter, and that the trapdoor problem is
already only subexponentially harder with it. 

\subsection{} 

Even if we take the private polynomial to be of higher Hamming weight, 
the amount of calculi required to recover it is almost the
same. Recall that the size of the $\mathcal{PK}$ is already almost
impractical. 

\subsection{} 

The problem of solving a single univariate pseudoquadratic polynomial
equation upon finite fields is an $\mathcal{NP}$-complete problem
\cite{kipnis99cryptanalysis}. So, it is reasonable to look for
cryptosystems that provide it as a trapdoor problem. The experience up
to now has shown that hiding polynomials does not help the security of
a cryptosystem, restricts choices, and renders the size of the
$\mathcal{PK}$ impractical. The privileged position of a legitimate
user must rely elsewhere.

\bibliographystyle{alpha}
\bibliography{biblio}
\end{document}